\documentclass[12pt]{article}
\usepackage{amssymb,amsmath}
\usepackage{epsfig}
\usepackage{epstopdf}
\usepackage{latexsym}
\usepackage{color}

\def\ds{\displaystyle}

\def\cF{{\cal F}}

\def\cN{{\cal N}}

\def\eql{~=~}
\def\Ga{{\Gamma}}
\def\mbM{\mathbb{M}}

\def\mbR{\mathbb{R}}
\definecolor{cardinal}{rgb}{0.6,0,0}
\definecolor{darkgreen}{rgb}{0,0.5,0}
\definecolor{golden}{rgb}{0.92, 0.7, 0}
\definecolor{midnight}{rgb}{0, 0, 0.5}
\definecolor{darkblue}{rgb}{0.2, 0, 0.8}

\newcommand{\Cardinal}{\color{cardinal}}

\def\r??{{\Cardinal ref?}}
\def\K{{\cal K}}
\def\kappa{{\cal K}}
\def\boxx{{\nabla^2_{X_5}}}
\def\boxxs{{\nabla^2_{}}}
\def\cF{{\cal F}}
\def\cA{{\cal A}}
\def\cK{{\cal K}}

\def\Phi{{\cal T}}

\begin{document}

\begin{titlepage}

\begin{flushright}
NSF-KITP-09-59
\end{flushright}

\bigskip
\bigskip
\bigskip
\centerline{\Large \bf Supersymmetric IIB Solutions  with
Schr\"{o}dinger Symmetry}
\bigskip
\bigskip
\bigskip
\centerline{{\bf Nikolay Bobev$^{1,2}$, Arnab Kundu$^2$ and
Krzysztof Pilch$^{2}$}}
\bigskip
\bigskip
\centerline{${}^{1}$ Kavli Institute for Theoretical Physics}
\centerline{University of California Santa Barbara}\centerline{
Santa Barbara, CA 93106-4030, USA}
\bigskip
\centerline{$^2$ Department of Physics and Astronomy}
\centerline{University of Southern California} \centerline{Los
Angeles, CA 90089, USA}
\bigskip
\centerline{{\rm bobev@usc.edu, akundu@usc.edu, pilch@usc.edu} }
\bigskip \bigskip

\begin{abstract}

We find a class of non-relativistic supersymmetric solutions of
IIB supergravity with non-trivial $B$-field that have dynamical
exponent $n=2$ and are invariant under the Schr\"{o}dinger group.
For a general Sasaki-Einstein internal manifold with $U(1)^3$
isometry, the solutions have two real supercharges. When the
internal manifold is $S^5$, the number of supercharges can be
four. We also find a large class of non-relativistic scale
invariant type IIB solutions with dynamical exponents different
from two. The explicit solutions and the values of the dynamical
exponents are determined by vector eigenfunctions and eigenvalues
of the Laplacian on an Einstein manifold.

\end{abstract}

\end{titlepage}

\section{Introduction}

The gauge/gravity duality has proven to be a powerful tool to
study strongly coupled field theories \cite{Aharony:1999ti}. There
are many strongly coupled condensed matter systems that are of
both theoretical and experimental interest. Thus it is reasonable
to ask how much can we learn about such field theories using the
AdS/CFT correspondence. There has been a lot of effort in this
direction as summarized recently in \cite{Hartnoll:2009sz,
Herzog:2009xv}.

There are non-relativistic condensed matter systems, like fermions
at unitarity, which exhibit the non-relativistic analog of the
conformal symmetry - the Schr\"{o}dinger symmetry. The
Schr\"{o}dinger algebra is generated by spatial translations
$P^i$, temporal translation $H$, spatial rotations $M^{ij}$,
Galilean boost $K^i$, the dilatation operator $D$, a special
conformal transformation $C$ and Galilean mass $M$. In a
non-relativistic scale invariant theory, time and space scale
differently, $t \to~\lambda^{n} t$ and $\vec{x}\to \lambda
\vec{x}$, respectively. The real parameter $n$ is called the
dynamical exponent and the Schr\"{o}dinger invariant systems have
$n=2$. When $n\neq 2$, one does not have the special conformal
generator in the algebra and the theory is only scale invariant.
The familiar case of scale invariance in a relativistic conformal
theory corresponds to $n=1$. More details  on non-relativistic
conformal theories can be found in
\cite{hagen,Mehen:1999nd,Nishida:2007pj}.

In \cite{Son:2008ye, Balasubramanian:2008dm}, a five-dimensional
gravitational background with Schr\"{o}dinger symmetry was found
as a solution to the Einstein-Hilbert action coupled  to a massive
vector field. Subsequently in \cite{Herzog:2008wg,
Maldacena:2008wh, Adams:2008wt} this solution was embedded in the
type IIB supergravity.\footnote{See
\cite{Duval:1990hj}-\cite{Pal:2009np} for examples of other
non-relativistic gravity solutions.}  The ten-dimensional
background was obtained by applying a solution generating
technique, known as the null Melvin twist, to the
  $AdS_5\times S^5$ background. The null Melvin twist,
described in \cite{Gimon:2003xk, Hubeny:2005qu}, can be used to
generate new  supergravity solutions starting   from a known
solution,  when the latter has at least one compact $U(1)$ and one
noncompact $U(1)$ isometry. This technique has been extensively
used in,   e.g., \cite{Mazzucato:2008tr, Bobev:2009zf} to generate
new non-relativistic gravity backgrounds.

There are also supersymmetric extensions of the Schr\"{o}dinger
algebra, which have been studied in \cite{beckers, gauntlett,
Leblanc:1992wu, Duval:1993hs, Henkel:2005dj,Sakaguchi:2008rx}.
Therefore it is natural to look for Schr\"{o}dinger invariant
supergravity solutions which   possess some supersymmetry. The
Schr\"{o}dinger invariant solutions discussed in
\cite{Herzog:2008wg, Maldacena:2008wh, Adams:2008wt} completely
break supersymmetry even though they are obtained from
supersymmetric IIB backgrounds. A supersymmetric Schr\"{o}dinger
invariant solution was constructed in \cite{Hartnoll:2008rs},
however, this solution has vanishing $B$-field and is sourced only
by the usual self-dual RR flux. Other examples of supersymmetric
non-relativistic solutions were found in \cite{Hartnoll:2008rs,
Donos:2009en, Colgain:2009wm, Pal:2009np}, however, all those
solutions have dynamical exponent $n\neq2$ and therefore are scale
invariant but not invariant under the full Schr\"{o}dinger group.

In this paper we will analyze in detail the supersymmetries
preserved by non-relativistic Schr\"{o}dinger invariant solutions
of the type IIB supergravity with non-vanishing $B$-field. We
consider non-relativistic backgrounds generated by  the null
Melvin twist applied  to the  Freund-Rubin type  solutions of the
form $AdS_5\times X_5$, where $X_5$ is a Sasaki-Einstein manifold.
When the Killing spinor of the Sasaki-Einstein manifold is
invariant under the $U(1)$ isometry used in the twist, the
non-relativistic solution preserves, in general, two real
supercharges. These two supercharges are a subset of the
Poincar\'{e} supersymmetries of the relativistic superconformal
algebra. The superconformal symmetries are completely broken by
the twist.

We illustrate our general result with two familiar examples: $S^5$
and $T^{1,1}$. In the first case, we find that the
non-relativistic solution can have four real supersymmetries. This
is due to $\cN=8$  unbroken supersymmetries in the original
solution on $S^5$ before the twist. The second case illustrates
better the generic situation where only $\cN=1$ supersymmetry is
present before the twist. Other examples that   are covered by our
   analysis are generalizations of the $T^{1,1}$ example and include   two
infinite families of Sasaki-Einstein manifolds, $Y^{p,q}$
\cite{Gauntlett:2004yd} and $L^{p,q,r}$ \cite{Cvetic:2005ft}. All
those spaces  can be used as internal manifolds for supersymmetric
Schr\"{o}dinger invariant IIB solutions. Since both of these
infinite families have $U(1)$ isometries that leave the Killing
spinor invariant, we find an infinite number of Schr\"{o}dinger
invariant solutions which preserve two supercharges.

The general form of the backgrounds constructed by the null Melvin
twist also suggests a natural Ansatz for non-relativistic type IIB
solutions with higher dynamical exponents and non-zero $B$-field.
We show that there is a large class of such solutions of the form
\begin{align*}
 ds_{10}^2 & \eql  -{\Omega\over   z^{2n}}du^2+{1\over z^2}(-2\,dudv+dx_1^2+dx_2^2+dz^2) +
 ds^2_{X_5}\,,\notag\\[6pt]
 F_{(5)} & \eql (1+\star)\,{\rm vol}_{X_5}\,,\\[6pt]
  B_{(2)} & \eql  {1\over z^{n}}\mathcal{A}\wedge du\,, \notag
\end{align*}
where $X_5$ is an Einstein manifold and $\mathcal{A}$ is an
one-form on $X_5$. We find that $\mathcal{A}$ must be a vector
eigenfunction of the Laplacian on $X_5$ and the dynamical
exponent, $n$, is determined by the corresponding eigenvalue. The
metric function $\Omega$ obeys an inhomogeneous scalar Laplace
equation on $X_5$. In principle, both $\cA$ and $\Omega$ can be
determined explicitly  using  harmonic expansions.

The   class of solutions constructed here includes all solutions
generated by the null Melvin twist and also the solutions with
general dynamical exponents and vanishing $B$-field found in
\cite{Hartnoll:2008rs}. It is worth emphasizing that in the more
general case of solutions with a nontrivial $B$-field,  the
dynamical exponent is related to the  eigenvalues of vector
harmonics on Einstein manifolds.

The paper is organized as follows: In Section 2, we present the
non-relativistic Schr\"{o}dinger invariant supergravity
backgrounds obtained by the null Melvin twist and recast them in a
form that is convenient for  analysis of unbroken supersymmetries
carried out in detail in Section 3. Then, in Section 4,  we work
out  some explicit examples that illustrate the general discussion
in Section 3. In Section 5, we introduce an Ansatz for type IIB
solutions with general dynamical exponents and show that it
reduces to a coupled system of a vector and a scalar Laplace
equations on the internal manifold. We also work out in detail
some examples on $S^5$ using standard methods of harmonic
expansion. We conclude in Section 6 with   comments and directions
for further study. A brief discussion of the null Melvin twist and
a summary of some pertinent solutions  are given in the Appendix.

\section{The solution}

Consider a Freund-Rubin type solution of IIB supergravity of the
form $AdS_5\times X_5$ with the metric and the five-form flux
given by
\begin{equation}
\label{frrub}
 ds^2_{10}\eql ds_{AdS_5}^2 + ds_{X_5}^2\,,
\end{equation}
\begin{equation}
\label{F5flux}
 F_{(5)}\eql (1+\star){\rm vol}_{AdS_5}\,,
\end{equation}
where $X_5$ is an Einstein manifold. In addition we assume that
$X_5$ has at least $U(1)$ isometry, with the corresponding Killing
vector $\K$.

In the following we will use  the metric on $AdS_5$ written in
terms of light-cone coordinates,
\begin{equation}
 ds_{AdS_5}^2\eql {1\over z^2}(-2\,dudv+dx_1^2+dx_2^2+dz^2)\,,
\end{equation}
with the radius of $AdS_5$ normalized to one.

The null Melvin twist \cite{Gimon:2003xk,Hubeny:2005qu} along a
Killing vector $\K$ on $X_5$ yields  another type IIB solution of
the form $Sch_5\times X_5$, where $Sch_5$ is a five-dimensional
space-time invariant under the Schr\"odinger
symmetry.\footnote{See the Appendix for more details of this
construction.} The metric,
\begin{equation}
\label{Schmet}
 ds_{10}^2\eql ds_{Sch_5}^2+ds_{X_5}^2\,,
\end{equation}
the five-form flux
\begin{equation}
\label{F5schflux}
 F_{(5)}\eql (1+\star){\rm vol}_{Sch_5}\,,
\end{equation}
and, in addition, a nonzero three-form flux, $H_{(3)}=dB_{(2)}$,
in the solution can be written explicitly    in terms of the data
of the  initial solution (\ref{frrub}), (\ref{F5flux}) and $\K$.
Specifically, the metric along $Sch_5$ in the  same light-cone
coordinates as above is
\begin{equation}
 ds_{Sch_5}^2 \eql -{\Omega\over   z^4}du^2+{1\over z^2}(-2\,dudv+dx_1^2+dx_2^2+dz^2)\,,
\end{equation}
and
\begin{equation}
\Omega=||\K||^2\,,
\end{equation}
is a nonnegative function given  by  the length square of the
Killing vector, $\K$, with respect to the metric on $X_5$.
Similarly, the two-form potential is given by
\begin{equation}
 \label{Bpot}
 B_{(2)}\eql {1\over z^2}\kappa\wedge du\,,
\end{equation}
where $\kappa$ is the one-form dual to $\K$.\footnote{In terms of
explicit coordinates $\xi^\alpha$ on $X_5$, we have $ds_{X_5}^2=
g_{\alpha\beta}d\xi^\alpha  d\xi^\beta$, $\Omega=
g_{\alpha\beta}\K^\alpha \K^\beta$, and
$\kappa_\alpha=g_{\alpha\beta}\K^\beta$.} To make sense of these
solutions as holographic duals to non-relativistic field theories
the light-cone coordinate $v$ should be periodically identified $v
\sim v + 2\pi r_{v} $
\cite{Son:2008ye,Balasubramanian:2008dm,Adams:2008wt}. The
momentum along this compact direction is quantized in units of the
inverse radius $r_v^{-1}$. This momentum is interpreted as the
Galilean mass (or the particle number) in the dual field theory.

By construction, (\ref{Schmet}), (\ref{F5schflux})  and
(\ref{Bpot}) satisfy the equations of motion\footnote{We have also
checked this explicitly, see the Appendix for more details.} of
type IIB supergravity for any Killing vector $\K$, since all one
is using is a series of boosts, T-dualities and shifts which are
all symmetries of IIB supergravity. One can also view the
$Sch_5\times X_5$ solution as a deformation of the $AdS_5\times
X_5$ solution above, which can be formally recovered by setting
$\K=0$ in (\ref{Schmet}), (\ref{F5schflux}) and (\ref{Bpot}).

It is possible that the norm of the Killing vector $\K$ vanishes
on some locus in $X_5$. The curvature of the solution is
completely regular on this locus, in fact, the solution looks like
$AdS_5\times X_5$. It is somewhat strange that the asymptotic
structure of the non-compact space changes from the
non-relativistic $Sch_5$ to $AdS_5$ at special points on $X_5$.
However this kind of space-times have been analyzed in the
literature, see \cite{Marolf:2002ye,Hubeny:2005qu} and references
therein. One can argue that, despite the presence of the locus on
which $\Omega$ vanishes, the ten-dimensional background is
non-distinguishing and thus has the proper asymptotic and causal
structure for a dual of a non-relativistic field theory. An
intuitive way to understand this is to observe that in the
solution (\ref{Schmet}) every point with $u>u_0$ can be reached by
a causal curve on the ten-dimensional background starting at
$u_0$. This implies that the light-cone is degenerate and the
space-time is non-distinguishing. This is precisely a property one
should expect from a gravity dual to a non-relativistic field
theory. The presence of the locus on which $\Omega$ vanishes does
not change the fact that the space-time is non-distinguishing as
long as $\Omega$ is non-zero on an open set in $X_5$. This will
always be the case for our solutions.\footnote{We are grateful to
Veronika Hubeny and Mukund Rangamani for helpful explanations on
this point.}

\section{General supersymmetry analysis}

Let us now assume that the $AdS_5\times X_5$ solution preserves
some of the supersymmetries of IIB supergravity, that is there
exists a chiral Killing spinor $\epsilon_0$ in ten dimensions,
\begin{equation}
 \label{chiral}
 \Gamma^1\ldots\Gamma^{10}\epsilon_0\eql\epsilon_0\,,
\end{equation}
for which  the dilatino and the gravitino supersymmetry variations
vanish. In the following we will find sufficient conditions under
which the Killing spinor, $\epsilon_0$, can be deformed to a
Killing spinor, $\epsilon$, of the $Sch_5\times X_5$ solution.

To this end let us  introduce the frames, $e^M$,  for the metric
(\ref{Schmet}),
\begin{align}
 \label{frames}
\notag  e^1 &\eql {1\over 2z^2}(\Omega+1)\,du+dv\,,\qquad  e^4 \eql {1\over 2z^2}(\Omega-1)\,du+dv\,, \\[6 pt]
  e^2 &\eql {1\over z}dx_1\,,\qquad e^3\eql {1\over z}dx_2\,,\qquad e^5 \eql {1\over z} dz\,,\\[6 pt]
 \notag  e^{5+\alpha} & \eql e_{(5)}^\alpha\,,\quad \alpha=1,\ldots,5\,.
\end{align}
where $e_{(5)}^\alpha$ are some orthonormal frames on $X_5$ that
will be specified later. The equations for unbroken supersymmetry
are \cite{Schwarz:1983qr}
\begin{equation}
\label{susydil}
 \delta\lambda\eql -{1\over 24} H_{MNP}\Gamma^{MNP}\epsilon^{*}\eql 0\,,
\end{equation}
and
\begin{equation}
\label{susypsi} \delta\psi_M\eql \nabla_M\epsilon+{i\over
480}F_{NPQRS}\Gamma^{NPQRS}\Gamma_M\,\epsilon-{1\over
48}\left[\Gamma_M,H_{PQR}\Gamma^{PQR}\right]\,\epsilon^*\,,
\end{equation}
where the flat indices  $M,N,\ldots$ range from 1 to 10. We use
the same conventions as in  \cite{Becker:2007zj} with the mostly
plus metric and real Dirac $\Gamma$-matrices in ten dimensions.

The corresponding frames, $e^M_0$,  and the supersymmetry
equations, $\delta_0\lambda=0$ and $\delta_0\psi_M=0$ for the
$AdS_5\times X_5$ solution are obtained by setting $\K=0$ in
(\ref{frames}),  (\ref{susydil}) and (\ref{susypsi}). In that case
the $H_{(3)}$ flux vanishes and the dilatino variation vanishes
identically.

We start our analysis of unbroken supersymmetries with the
dilatino variation (\ref{susydil}) in which the   $H_{(3)}$ flux
is given by
\begin{align}
  H_{(3)} & \eql dB_{(2)} \eql \left(d\kappa +2\kappa\wedge
e^5\right)\wedge (e^1-e^4)\,.
\end{align}
Since $\kappa$ is a one form on $X_5$, it follows that
(\ref{susydil}) factorizes into
\begin{equation}
\label{facla} \delta\lambda \eql -{1\over
24}H_{MNP}\Gamma^{MNP}\epsilon^*\eql
\mbM(\Gamma^1-\Gamma^4)\epsilon^* \eql 0\,,
\end{equation}
where $\mbM$ is a real matrix
\begin{equation}
\label{Mmat}
 \mbM\eql -{1\over 8} (d\kappa)_{MN}\Gamma^{MN}+{1\over 2}\kappa_M\Gamma^{5M}\,.
\end{equation}
Note that the summation above is over the range $M,N=6,\ldots,10$
since  both $\kappa$ and $d\kappa$ have  nonvanishing components
only along $X_5$. Hence we can solve (\ref{facla}) by  imposing a
single  projection condition
\begin{equation}
 \label{proj14}
 (\Gamma^1-\Gamma^4)\epsilon^*\eql  (\Gamma^1-\Gamma^4)\epsilon\eql 0\,,
\end{equation}
where the condition for $\epsilon$ follows using   reality of the
$\Gamma$-matrices.

The gravitino variations (\ref{susypsi}) involve two types of
terms that depend on $\K$ and thus are absent in the corresponding
equations for $\epsilon_0$. If those terms can be eliminated from
the equations, the problem of finding the Killing spinor
$\epsilon$ on $Sch_5\times X_5$ will be reduced to that of finding
$\epsilon_0$ on $AdS_5\times X_5$.

The additional terms of the first type arise from the deformation
of the spin connection due to the function $\Omega$ in $e^1$ and
$e^4$. We can write this deformation succinctly as the difference
of the spin connections for the two metrics,
\begin{align}
\label{tors}
 W-W\big|_{\Omega=0}\eql {4\Omega\over z^{5}}\,du\otimes du\wedge dz
-{1\over z^{4}}\,du\otimes du\wedge d\Omega\,,
\end{align}
where
\begin{equation}
 W\eql \omega_{MN}\otimes e^M\wedge e^N\,,
\end{equation}
and $\omega_{MN}$ is the spin connection. It is clear that the
deformation due to those additional terms will arise only in the
$\delta\psi_u$ variation.\footnote{We use a shorthand notation for
the curved indices labelling them with the corresponding
coordinate. }

The deformation terms involving the $H_{(3)}$ flux manifestly
vanish due to (\ref{proj14}) for all $\delta\psi_M$, but
$\delta\psi_1$ and $\delta\psi_4$. Indeed, for $M\not=1,4$, the
$(\Gamma^1-\Gamma^4)$ factor arising from the contraction  as in
(\ref{facla}) commutes, or anticommutes,  with all other matrices
in this term. Hence it can be moved to act directly on
$\epsilon^*$, so that these variations vanish due to
(\ref{proj14}).

To evaluate the remaining two variations, consider the combination
$e^1\delta\psi_1+e^4\delta\psi_4$. This yields a sum of two terms
\begin{equation}
\label{hhs}
 {1\over 96 } \left({\Omega\over z^2} du+2 \,dv\right)\left[\Gamma^1-\Gamma^4,H_{MNP}\Gamma^{MNP}\right]\epsilon^*+{1\over 96z^2}du\left[\Gamma^1+\Gamma^4,H_{MNP}\Gamma^{MNP}\right]\epsilon^*\,.
\end{equation}
The first commutator vanishes identically, while the second one
gives a nontrivial contribution to $\delta\psi_u$, which we
evaluate explicitly below.

To summarize, we have shown that, apart from the dilatino
variation, the only equation that is modified by the deformation
is the gravitino variation $\delta\psi_u$. Before we proceed with
this variation, let us note that the  other gravitino variations
along $Sch_5$ are solved by a single additional projector,
\begin{equation}
 \label{proj23}
 \left(\Gamma^2+i\Gamma^3\right)\epsilon\eql 0\,.
\end{equation}
Indeed, upon using (\ref{chiral}) to simplify the $F_{(5)}$ flux
terms, and then imposing the projection (\ref{proj14}), all
variations $\delta\psi_{x_1}$, $\delta\psi_{x_2}$, $\delta\psi_v$
and $\delta\psi_z$ reduce to (\ref{proj23}) multiplied by some
other $\Gamma$-matrix.

Note that in the case of the $AdS_5\times X_5$ solution, the
gravitino variations along $AdS_5$ are solved by a single
projector
\begin{equation}
 \label{proj1234}
 \left(1-i\Ga^{1234}\right)\epsilon\eql 0\,.
\end{equation}
The solutions to this equation include both solutions to
(\ref{proj14}) and (\ref{proj23})  and to the equations where both
projectors are replaced by the ones with the opposite sign.

Finally, consider the variation $\delta\psi_u$. Here we find
\begin{align}
\label{uvar}
\notag  \delta\psi_u-\delta_0\psi_u & \eql -{1\over 4 z^2}\left(3\,\Omega \,\Ga^5 -\Gamma^\alpha \partial_\alpha\Omega\right)(\Ga^1-\Ga^4)\,\epsilon\\[6 pt]
&\qquad +{i\over 4z^2}\,\Omega\,\Ga^2\Ga^3\Ga^5(\Ga^1-\Ga^4)\,\epsilon\\[6 pt]
\notag  &\qquad -{1\over  z^2}\,\mbM\,\epsilon^*\,.
\end{align}
The terms in the first line are due to (\ref{tors}), and we have
introduced a shorthand notation
$\Gamma^\alpha=e^\alpha_M\Gamma^M$. The second line arises from
additional terms in the $F_{(5)}$ flux in the coordinate basis due
to the $\Omega$-terms in $e^1$ and $e^4$. The last line is due to
the non-vanishing term in (\ref{hhs}) with the matrix $\mbM$ given
in (\ref{Mmat}). Clearly the first two lines vanish if we impose
(\ref{proj14}), which leaves a single additional algebraic
constraint on the Killing spinor,
\begin{equation}
 \label{Meqs}
 \mbM\,\epsilon^{*}\eql 0\,.
\end{equation}
In the following we will unravel the conditions under which this
equation has nontrivial solutions.

The transformation of a  Killing spinor, $\epsilon$, under an
isometry $\K$ is given by the Lie derivative
\begin{align}
 \label{lieder}
\notag  {\cal L}_\K\,\epsilon & \eql \K^M\partial_M\epsilon+{1\over 4}\left(\K^M\omega_{MPQ}+\nabla_{[P}\K_{Q]}\right)\Gamma^{PQ}\epsilon\\[6 pt]
 & \eql \K^M\nabla_M\epsilon+ {1\over 8}(d\kappa)_{MN}\Gamma^{MN}\epsilon\,.
\end{align}
Next consider the gravitino variation along $\K$,
\begin{equation}
\label{killvar}
 \K^M\delta\psi_M\eql \K^M\nabla_M\epsilon+{i\over 480}F_{NPQRS}\Gamma^{NPQRS}(\K^M\Gamma_M)\,\epsilon\,.
\end{equation}
The second term can be expanded using the explicit form of the
$F_{(5)}$ flux in  (\ref{F5schflux}). We get, using (\ref{chiral})
and (\ref{proj1234}),
\begin{align}
\notag {i\over 480}F_{NPQRS}\Gamma^{NPQRS}(\K^M\Gamma_M)\,\epsilon
&  \eql {1\over 2}\kappa_M\Gamma^{5M}\epsilon\,.
\end{align}
Substituting this back in (\ref{killvar}) and using (\ref{lieder})
and (\ref{Mmat}), we obtain
\begin{align}
\notag   \K^M\delta\psi_M & \eql  {\cal L}_\K\,\epsilon -{1\over 8}(d\kappa)_{MN}\Gamma^{MN}\epsilon+{1\over 2}\kappa_M\Gamma^{5M}\epsilon\\[6 pt]
& \eql {\cal L}_\K\,\epsilon+\mbM\,\epsilon\,.
\end{align}
This shows that a Killing spinor, $\epsilon$, is annihilated by
$\mbM$ if and only if it is invariant under the corresponding
isometry. Since $\mathbb{M}$ is real, this also implies that
$\mathbb{M}~\epsilon^{*}=0$.

The result of our analysis is an explicit method  for obtaining
Killing spinors for the Schr\"odinger background, $Sch_5\times
X_5$ obtained by the null Melvin twist, starting with the Killing
spinors of the undeformed $AdS_5\times X_5$ background:

\begin{itemize}
{\it
\item A Killing spinor, $\epsilon$, on $AdS_5\times X_5$  is also a Killing spinor on  $Sch_5\times X_5$, where  the $Sch_5\times X_5$ solution is obtained by the null Melvin twist along  the Killing vector $\K$, provided $\epsilon$ satisfies
\begin{equation*}
{1\over 2}\left(1+\Gamma^{14}\right)\,\epsilon\eql 0 \qquad {\rm
and}\qquad {\cal L}_\K\,\epsilon\eql 0\,.
\end{equation*}
\item Conversely, any Killing spinor on $Sch_5\times X_5$ satisfying the $\Gamma^{14}$ projection above gives rise to a $\K$-invariant Killing spinor on $AdS_5\times X_5$.

}
\end{itemize}

In fact, it appears that the above construction gives rise to {\it
all}\,\, Killing spinors on $Sch_5\times X_5$. The complete
analysis is more involved. If we start with an  $\epsilon$ that
does not satisfy (\ref{proj14}), we must solve the dilatino
variation by setting $\mbM\,\epsilon^*=0$ from the start.
Furthermore, in all gravitino variations, the $H_{(3)}$ flux terms
will not cancel. A systematic method to exclude this type of
Killing spinors would be to analyze the integrability conditions
for the gravitino variations. We have not carried out this
calculation in the general case, but rather verified explicitly in
the simplest examples of $X_5=S^5$ and $X_5=T^{1,1}$ that there
are no further Killing spinors of opposite $\Gamma^{14}$
chirality. This is in agreement with
\cite{Hartnoll:2008rs,Donos:2009en}, where it was shown that
non-relativistic supersymmetric solutions with vanishing $H_{(3)}$
flux and different dynamical exponents break all superconformal
Killing supersymmetries.

The undeformed $AdS_5\times X_5$ solution has $4\otimes2 = 8$ real
supercharges where the factor of $4$ in the direct product comes
from the $AdS_5$ Killing spinors and $2$ is the number of Killing
spinors on a generic Sasaki-Einstein manifold. As discussed above,
the null Melvin twist breaks all $AdS_5$ supersymmetries and we
are left with non-relativistic solutions preserving $2$
supercharges. For the case of $AdS_5\times S^5$ we have $4 \otimes
8 = 32$ supersymmetries because $S^5$ has $8$ Killings spinors and
as we discuss in the next section one can find cases in which the
number of supersymmetries of the twisted solution is enhanced to
$4$.

\section{Examples}

In this section we illustrate how the general construction in the
previous section works for some well known five-dimensional
Sasaki-Einstein manifolds.

\subsection{$S^5$}

The most symmetric example of a five-dimensional Sasaki-Einstein
manifold is the sphere $S^5$. It has $SO(6)$ isometry group and we
can apply the general, three-parameter null Melvin twist on a
$U(1)^3$ subgroup. One can find the conditions for unbroken
supersymmetry and construct Killing spinors of the twisted
solution explicitly, however, it is more efficient to use group
theory to extract this information.

The $SO(6)$ isometry group of $S^5$ is generated by the Killing
vectors $M_{IJ}=x_I\partial_J-x_J\partial_I$, where we realize
$S^5$ as a unit sphere $x_1^2+\ldots +x_6^2=1$ in $\mbR^6$. If we
choose the $U(1)^3$ Cartan subalgebra generators as
\begin{equation}
 \K_{(1)}\eql M_{12}\,,\qquad \K_{(2)}\eql M_{34}\,,\qquad \K_{(3)}\eql M_{56}\,,
\end{equation}
we find that for a Killing vector
$\K=\eta_1\K_1+\eta_2\K_2+\eta_3\K_3$,
\begin{equation}
 \Omega\eql ||\K||^2\eql \eta_1^2(x_1^2+x_2^2)+\eta_2^2(x_3^2+x_4^2)+\eta_3^2(x_5^2+x_6^2)\,.
\end{equation}
The Killing spinors on $S^5$ transform in ${\bf 4}\oplus\bar{\bf
4}$ of $SO(6)$. Their charges with respect to the $U(1)^3$ above
are
\begin{equation}
 {\bf 4}\quad \longrightarrow\quad (+1,+1,+1)\oplus (+1,-1,-1)\oplus(-1,+1,-1)\oplus(-1,-1,+1)\,.
\end{equation}
Hence the Killing spinors invariant under $\K$ are determined by
solutions to the equation
\begin{equation}
 \eta_1\pm\eta_2\pm\eta_3\eql 0\,. \label{eta_iS5}
\end{equation}
For values of $\eta_i$ satisfying (\ref{eta_iS5}), the
non-relativistic solution preserves two real supersymmetries. If
in addition to (\ref{eta_iS5}) we impose that at least one of the
$\eta_i$ vanishes, the number of real supercharges is enhanced to
four. One can verify explicitly that all sixteen superconformal
Killing spinors of $AdS_5\times S^5$ are broken by the null Melvin
twist so that one can preserve only a subset of the Poincar\'{e}
Killing spinors. Of course when all $\eta_i$ vanish we get back to
the original $AdS_5\times S^5$ background, which preserves sixteen
superconformal and sixteen Poincar\'{e} supersymmetries.

It is clear that $\Omega$  is strictly positive when none of the
$\eta_i$'s vanish. Setting one $\eta_i$ to zero, say $\eta_3=0$,
the Killing vector $\K$ vanishes on $S^1$ given by
$x_1=x_2=x_3=x_4=0$ and $x_5^2+x_6^2=1$. Similarly, when two
$\eta_i$ vanish, there is an $S^3$ on which $\K$ vanishes. As we
discussed in Section~2, even though there could be a locus on
which $\Omega$ vanishes, the twisted background is still
non-distinguishing and thus non-relativistic.

The special case $\eta_1=\eta_2=\eta_3= \eta$ corresponds to the
null Melvin twist along the Hopf fiber of $S^5$ and has been
studied in \cite{Herzog:2008wg,Maldacena:2008wh,Adams:2008wt}. In
this case the Killing vector $\K$ has constant norm and it is
clear from (\ref{eta_iS5}) that the twisted solution breaks
supersymmetry completely. This has been also shown explicitly in
\cite{Maldacena:2008wh}.

\subsection{$T^{1,1}$}

In this section we describe explicitly the null Melvin twists for
the $AdS_5\times T^{1,1}$ solution \cite{Romans:1984an}. Recall
that $T^{1,1}$ is the coset space  $SU(2)\times SU(2)/U(1)$ with a
unique homogenous Einstein metric. There is a Killing spinor on
$T^{1,1}$, with two real components, which gives rise to the
$\mathcal{N}=1$ unbroken supersymmetry of the Romans solution. The
isometries of the solution arise from the obvious $SU(2)\times
SU(2)$  action on the left and, in addition, from another  $U(1)$
action from the right on the coset. The Killing spinor is
necessarily invariant under $SU(2)\times SU(2)$, and transforms
nontrivially under the $U(1)$, which is the $R$ symmetry of the
$\mathcal{N}=1$ superalgebra.

It is convenient to realize $T^{1,1}$ explicitly as a locus in
$\mathbb{C}^4$ by  introducing a complex matrix
\cite{Candelas:1989js}
\begin{equation}\label{consr}
 {\cal W}\eql {1\over \sqrt 2}\left(
 \begin{matrix}
 z_3+iz_4 & z_1-iz_2 \\
  z_1+iz_2 & -z_3+i z_4
 \end{matrix}
  \right)\,,
\end{equation}
subject to the constraints
\begin{equation}
 \mathop{{\rm Tr}}{\cal W}^\dagger{\cal W}=1\,,\qquad \det{\cal W}\eql 0\,.
\end{equation}
In this parametrization,  the two $SU(2)$'s, call them $SU(2)_1$
and $SU(2)_2$, act on $\cal W$ by left and right multiplication,
respectively, while the $R$-symmetry, $U(1)_R$,  corresponds to
the phase rotation, $z_i\rightarrow e^{i\phi_3}z_i$. As explained
in \cite{Candelas:1989js}, the constraints (\ref{consr}) can be
solved explicitly by introducing the Euler angles,
$(\theta_1,\phi_1,\psi_1)$ and $(\theta_2,\phi_2,\psi_2)$ for
$SU(2)_1$ and $SU(2)_2$, and setting $\psi_1=\psi_2=\phi_3/2$ to
pass onto the coset. In terms of those angles, the unique Einstein
metric on $T^{1,1}$ is \cite{Candelas:1989js}
\begin{multline}
ds^2_{T^{1,1}} = \ds\frac{1}{6} ( d\theta_1^2 + \sin^2\theta_1
d\phi_1^2+d\theta_2^2 + \sin^2\theta_2d\phi_2^2 )+
\ds\frac{1}{9}(d\phi_3+\cos\theta_1d\phi_1+\cos\theta_2d\phi_2)^2\,.
\end{multline}
For the general null Melvin twist we choose Killing vectors
$\K_{(1)}$, $\K_{(2)}$ and $\K_{(3)}$ corresponding to
$U(1)_1\subset SU(2)_1$, $U(1)_2\subset SU(2)_2$ and $U(1)_R$,
normalized such that
\begin{equation}
\K_{(1)}\eql{\partial\over\partial\phi_1}\,,\qquad
\K_{(2)}\eql{\partial\over\partial\phi_2}\,,\qquad
\K_{(3)}\eql{\partial\over\partial\phi_3}\,.
\end{equation}
One can perform the null Melvin twist along $\phi_i$ and the
Killing vector defining the non-relativistic solution is
\begin{equation}
\K = \ds\sum_{i=1}^3 \eta_i \ds\frac{\partial}{\partial\phi_i}~.
\end{equation}
The function $\Omega$ is
\begin{multline}
\Omega = \eta_1^2 \left( \ds\frac{\sin^2\theta_1}{6} +
\ds\frac{\cos^2\theta_1}{9} \right) + \eta_2^2 \left(
\ds\frac{\sin^2\theta_2}{6} + \ds\frac{\cos^2\theta_2}{9} \right)
+ \ds\frac{\eta_3^2}{9} \\+ 2\eta_1\eta_2
\ds\frac{\cos\theta_1\cos\theta_2}{9} + 2\eta_1\eta_3
\ds\frac{\cos\theta_1}{9} + 2\eta_2\eta_3
\ds\frac{\cos\theta_2}{9}\ .
\end{multline}
The matrix $\mathbb{M}$ is given by
\begin{eqnarray}
\mathbb{M}  &=& \ds\frac{\eta_1}{2}
\left(\ds\frac{\sin\theta_1}{\sqrt{6}}
 \left( \Gamma^{58} + \Gamma^{610}\right) - \ds\frac{\cos\theta_1}{3} (2\Gamma^{68}-\Gamma^{79} -
\Gamma^{510}) \right)
\\&&+~\ds\frac{\eta_2}{2}
\left(\ds\frac{\sin\theta_2}{\sqrt{6}}
 \left( \Gamma^{59} + \Gamma^{710}\right) - \ds\frac{\cos\theta_2}{3} (2\Gamma^{79}-\Gamma^{68} -
\Gamma^{510}) \right) + \ds\frac{\eta_3}{6}
(\Gamma^{68}+\Gamma^{79} + \Gamma^{510})\ . \notag
\end{eqnarray}
One can show that the Killing spinor for these solutions is
\begin{equation}
\epsilon \eql e^{-\frac{i}{2}\phi_3}\widetilde{\epsilon}_0\
,\label{spinorT11}
\end{equation}
where $\widetilde{\epsilon}_0$ is a constant spinor satisfying the
chirality condition (\ref{chiral}) and four additional projectors
\begin{equation}
(1+\Gamma^{14})\widetilde{\epsilon}_0 \eql
(1+i\Gamma^{23})\widetilde{\epsilon}_0 \eql (1+i
\Gamma^{68})\widetilde{\epsilon}_0 \eql
(1+i\Gamma^{79})\widetilde{\epsilon}_0 \eql 0 \,.
\end{equation}
The condition $\mathbb{M}~\epsilon^{*}=0$ is satisfied by
(\ref{spinorT11}) if $\eta_3=0$.

Thus we find that the generalized null Melvin twist of
$AdS_5\times T^{1,1}$ with the parameters $(\eta_1,\eta_2,0)$ for
non-zero $(\eta_1,\eta_2)$ leads to a type IIB solution of the
form $Sch_5\times T^{1,1}$ with $H_{(3)}$ and $F_{(5)}$ flux,
which preserves two real supercharges. For $\eta_3\neq0$, we still
have a Schr\"{o}dinger invariant type IIB solution, but the
supersymmetry is completely broken.

\subsection{$Y^{p,q}$ and $L^{p,q,r}$}

There are two infinite families of five-dimensional
Sasaki-Einstein manifolds with explicitly known metrics. The
manifolds in the $Y^{p,q}$ family, found in
\cite{Gauntlett:2004yd}, are specified by two integers $(p,q)$
determined by some regularity conditions.\footnote{For our
purposes we will not distinguish between regular and quasi-regular
Sasaki-Einstein manifolds \cite{Gauntlett:2004yd}.} The solutions
have $SU(2)\times U(1)\times U(1)_R$ symmetry. The $L^{p,q,r}$
solutions are specified by a set of integers $(p,q,r)$ and have
even smaller isometry group, $U(1)^2\times U(1)_R$
\cite{Cvetic:2005ft}. The $U(1)_R$ isometry is special and the
Killing vector\footnote{Also called the Reeb vector.}
corresponding to it has a constant norm. Such a Killing vector
exists on every Sasaki-Einstein manifold, $X_5$, and it can be
determined by the K\"{a}hler form on the corresponding Calabi-Yau
cone over $X_5$ \cite{Gibbons:2002th}. The Killing spinor on $X_5$
has two real components and is charged under $U(1)_R$ so the
results of Section 3 imply that the null Melvin twist along
$U(1)_R$ will break supersymmetry completely. However, for both
the $Y^{p,q}$ and the $L^{p,q,r}$ families we have two additional
$U(1)$ isometries along which we can apply the twist with
arbitrary real parameters $(\eta_1,\eta_2)$. The resulting
non-relativistic solutions will be Schr\"{o}dinger invariant and
will preserve two real supercharges that are a subset of the
Poincar\'{e} supersymmetries of $AdS_5\times X_5$. The
superconformal charges are completely broken by the twist. Since
the metrics on both $Y^{p,q}$ and $L^{p,q,r}$ are explicitly
known, one can in principle construct explicit Killing spinors on
them.

\section{New solutions from vector harmonics}

In this section we introduce a new class of solutions with
Galilean symmetry, general dynamical exponents and nontrivial
three-form flux that are  generated by vector harmonics  on $X_5$.
The starting point of our construction is an Ansatz that is a
natural generalization of the twisted solutions in Sections 2 and
3 and the   solutions with general dynamical exponents, but
without $H_{(3)}$ flux, constructed recently by  Hartnoll and
Yoshida \cite{Hartnoll:2008rs} using scalar harmonics on $X_5$.

In the notation of Section 2, the metric in \cite{Hartnoll:2008rs}
is of the form
\begin{equation}
 \label{eqn:Hart}
 ds^2 \eql -{\Omega\over   z^{2n_1}}du^2+{1\over z^2}(-2\,dudv+dx_1^2+dx_2^2+dz^2)+ds_{X_5}^2\,,
\end{equation}
where $\Omega$ is a function on an internal Einstein manifold,
$X_5$, and $n_1$ is a real positive constant. The five form flux
remains the same, $F_{(5)}=(1+\star)\,{\rm vol}_{X_5}$. We
complete the Ansatz by introducing a   three form flux with the
potential
\begin{equation}\label{eqn:flux}
 B_{(2)}\eql {1\over z^{n_2}} {\cal A}\wedge du\,,\qquad H_{(3)}\eql dB_{(2)}\,,
\end{equation}
where $\cal A$ is an arbitrary one-form on $X_5$ and $n_2$ is a
real constant.

The   type IIB field equations \cite{Schwarz:1983qr} for this set
of fields read
\begin{equation}
 \label{eqn:sugein}
 R_{MN}\eql \frac{1}{6} F_{MPQRS} F_N{}^{PQRS} +\frac{1}{4}H_{MPQ}H_N{}^{PQ}  \,,
\end{equation}
and, taking into account the form of the $H_{(3)}$ flux with
nonzero components only along mixed directions,
\begin{equation}
 \label{eqn:maxwell}
 \nabla^MH_{MNP}\eql 0\,.
\end{equation}

The only nonvanishing component of the Einstein equations
(\ref{eqn:sugein}) is along the $uu$-direction where it reduces to
\begin{equation}
 \label{eqn:redeein}
 {1\over z^{2n_1}}\,\left({1\over 2}\,\boxx\,\Omega +2\,(n_1^2+1)\Omega\right) -{4\over z^{2n_1}}\,\Omega \eql{1\over z^{2 n_2}}\,\left({1\over 4}\,{\cal F}_{\alpha\beta}{\cal F}^{\alpha\beta}+{n_2^2\over 2}\,{\cal A}_\alpha{\cal A}^\alpha\right) \,.
\end{equation}
The terms in the bracket on the left hand side  arise from the
Ricci tensor. The second term comes from the energy momentum
tensor of the five-form flux, where   the $\Omega$ dependence is
introduced by the vielbein $e_u{}^M$, see (\ref{frames}). Finally,
the right hand side  arises from the energy momentum tensor of the
three-form flux, where $\cF=d\cA$.

The Maxwell equations (\ref{eqn:maxwell}) reduce to two equations.
The component of (\ref{eqn:maxwell}) along $du\wedge dz$ gives
\begin{equation}
 \label{eqn:tran}
 n_2\nabla^\alpha\cA_\alpha\eql 0\,.
\end{equation}
The remaining components yield the covariant massive Proca
equation
\begin{equation}
 \label{eqn:Proca}
 \nabla^\alpha \cF_{\alpha\beta}+(n_2^2+2n_2)\,\cA_\beta\eql 0\,.
\end{equation}
Expressing components of $\cF$ in a covariant form,
 \begin{equation}
 \label{eqn:ftens}
 \cF_{\alpha\beta}\eql \nabla_\alpha\cA_\beta-\nabla_\beta\cA_\alpha\,,
\end{equation}
and using the transversality condition (\ref{eqn:tran}), and
$R_{\alpha\beta}=4g_{\alpha\beta}$, the latter equation can be
rewritten as\footnote{The operator $(\boxx-4)$ is, in our
normalization, the Lichnerowicz operator on vector fields on the
Einstein manifold $X_5$.}
\begin{equation}
 \label{eqn:veclap}
( \boxx -4)\cA_\alpha+(n_2^2+2n_2)\,\cA_\alpha\eql 0\,,
\end{equation}
which is the covariant Laplace equation for vector fields on
$X_5$.

\vspace{6 pt} Let us first discuss the solutions of
(\ref{eqn:redeein}) and (\ref{eqn:veclap}) in the known cases.

\begin{itemize}
\item $\cA_\alpha=0$\\[6pt]
We can solve Maxwell equation  (\ref{eqn:veclap}) by setting
$\cA_\alpha=0$. In this case one is left with a Laplace equation
\begin{equation}
 \boxx\Omega+4(n^2-1)\,\Omega\eql 0\,,
\end{equation}
on the scalar harmonics on the Einstein manifold, $X_5$, where we
set $n=n_1$. This  case has been discussed in detail in
\cite{Hartnoll:2008rs}. Here we only note that the discrete
eigenvalues of the Laplacian determine the  discrete set of
dynamical exponents $n$. The specific values depend of course on
the choice of $X_5$.
\end{itemize}

When $\cA_\alpha$ does not vanish, the two exponents must be
equal, $n_1=n_2=n$. Indeed, since both terms on the right hand
side in (\ref{eqn:redeein}) are manifestly positive, the powers of
$z$ on both sides of the equation must be the same.

\begin{itemize}

\item $\cA_\alpha=\cK_\alpha$ is a Killing vector\\[6 pt]
In this case we can use the standard fact that on an Einstein
manifold Killing vectors are eigenfunctions of the Laplacian.
Since we choose normalizations such that the internal metric is of
unit radius, we have,\footnote{See, (\ref{eqn:killderLap}) in the
Appendix.}
\begin{equation}
 \label{laponkill}
 \boxx\,\K_\alpha\eql -4\,\K_\alpha\,.
\end{equation}
This   solves (\ref{eqn:veclap}) for $n=2$. For this value of $n$,
all terms without derivatives in (\ref{eqn:redeein}) cancel if we
take $\Omega=\K_\alpha\K^\alpha$. Then the derivative terms
combine into
\begin{equation}
  \K^\alpha\boxx\,\K_\alpha+{1\over 2}\left(\nabla_\alpha\K_\beta+\nabla_\beta\K_\alpha\right)\nabla^\alpha\K^\beta+4\,\K_\alpha\K^\alpha\eql 0\,,
\end{equation}
which obviously reduces to     (\ref{laponkill}). This verifies
explicitly that the backgrounds obtained by the null Melvin twist
along a Killing vector solve  the type IIB equations of motion.
\end{itemize}

In the general case,  we have a coupled system of Laplace
equations for the vector field, $\cA_\alpha$, and the function,
$\Omega$, on $X_5$,
\begin{align}
\label{eqn:laponvectors}
(\boxx-4)\,\cA_\alpha &\eql -n(n+2)\,\cA_\alpha\,,\\[6 pt]
\label{eqn:laponscalars} \boxx\Omega + 4(n^2-1)\,\Omega & \eql
\Phi(\cA)\,,
\end{align}
where
\begin{equation}\label{eqn:phifnct}
 \Phi(\cA)\eql {1\over 2}\,{\cal F}_{\alpha\beta}{\cal F}^{\alpha\beta}+{n^2}\,{\cal A}_\alpha{\cal A}^\alpha\,,
\end{equation}
is a scalar function on $X_5$.

It is clear that, at least in principle,  the system
(\ref{eqn:laponvectors}-\ref{eqn:phifnct}) can be solved
systematically using harmonic analysis on $X_5$. In the first step
one determines the spectrum of the operator $(\boxx-4)$ on vector
fields, which in turn determines the values of allowed
exponents~$n$ in (\ref{eqn:laponvectors}). For a given eigenvalue
of the vector Laplacian there is a degeneracy in the spectrum of
vector harmonics. This degeneracy depends on $X_5$ and will lead
to a family of solutions for a fixed eigenvalue. Next, for a given
$n$, one solves (\ref{eqn:laponvectors}) by setting $\cA_\alpha$
to be one of the vector harmonics for the corresponding
eigenvalue. The scalar function $\Phi(\cA)$ becomes then a source
for the inhomogeneous massive Laplace equation
(\ref{eqn:laponscalars}) for the function $\Omega$, which may be
solved by expanding $\Phi(\cA)$ into scalar spherical harmonics.
We will illustrate this procedure below by explicitly working out
some solutions for $X_5=S^5$ and by mapping out the relation
between vector eigenvalues of the Laplacian on $T^{1,1}$ and
dynamical exponents.

It is worth emphasizing that, for a given vector harmonic
$\cA_\alpha$, the scalar function  $\Omega$ in
(\ref{eqn:laponscalars}) may not be unique. This happens when
$-4(n^2-1)$, as determined by the eigenvalue of the vector
harmonic, $\cA_{\alpha}$, is an eigenvalue of the scalar
Laplacian. Then one can add to $\Omega$ a solution $\Omega_0$ of
the homogenous equation
\begin{equation}\label{eqn:homogen}
\boxx\Omega_0 + 4(n^2-1)\,\Omega_0=0\,.
\end{equation}
 In the next subsection we will see an
example of such non-uniqueness in the case when $X_5=S^5$.

A potential problem with a  general solution $(\cA_\alpha,\Omega)$
of (\ref{eqn:laponvectors}-\ref{eqn:phifnct}) is that there will
be regions in $X_5$ where the function $\Omega$ becomes negative.
This will change the causal and asymptotic structure of the
ten-dimensional solution and may lead to instabilities
\cite{Hartnoll:2008rs}. Clearly, it would be interesting to
understand properties of such solutions in more detail and, in
particular, to analyze their role, if any, in non-relativistic
holography.

Finally, let us note that some of the solutions with $n\neq2$  and
Sasaki-Einstein manifold $X_5$ may preserve some supersymmetry. As
we discussed in   Section 3, if the function $\Omega$ is non-zero,
the superconformal symmetries are broken, but some of the Killing
spinors on $X_5$ may be preserved. In the special case when
$\cA=0$ and $X_5$ is a Sasaki-Einstein manifold, one can use the
supersymmetry variations in Section 3 to show\footnote{This has
also been shown in \cite{Hartnoll:2008rs}.} that the
ten-dimensional solutions preserve two supercharges. For
$X_5=S^5$, the number of supercharges increases to  eight. We have
not performed a general analysis of the Killing spinor equations
for $n\neq2$ and $\cA\neq0$, but it would be interesting if some
of them turned out to be supersymmetric.

\subsection{Vector harmonics on $S^5$}

The scalar and vector harmonics on spheres  were extensively
discussed in the literature in the context of the Kaluza-Klein
reduction of supergravities.\footnote{For a recent comprehensive
review  in the present context the reader may consult
\cite{PvanNreview:2008}. We thank Peter~van~Nieuwenhuizen for
making this article available to us before publication.}   The few
basic facts that we need here are derived in
\cite{PvanNreview:2008} and \cite{Castellani:1991et}, where also
earlier references can be found.

All scalar and vector harmonics on $S^5$ can be constructed  from
the basic scalar harmonic $Y^A$ and the basic vector harmonic
$Y_\alpha^A$ that transform in the vector representation of
$SO(6)$ with components labeled by the index $A=1,\ldots,6$. They
satisfy the following algebraic constraints
\begin{equation}
 \label{eqn:algcon}
 \sum_{\alpha=1}^{5} Y_\alpha^A Y_\alpha^B\eql-Y^AY^B+\delta^{AB}\,,\qquad \sum_{A=1}^{6} Y^AY^A\eql 1\,,
\end{equation}
and form a closed system under covariant differentiation
\begin{equation}\label{eqn:derony}
 \nabla_\alpha Y^A\eql Y_\alpha^A\,,\qquad \nabla_\alpha Y^A_\beta\eql -\delta_{\alpha\beta}Y^A\,.
 \end{equation}
All   scalar harmonics are labeled by the totally symmetric
traceless representations of $SO(6)$ and are given by
\begin{equation}
 \label{scalarhar}
 Y^{A_1\ldots A_k}\eql Y^{(A_1}Y^{A_2}\ldots Y^{A_k)}\,,\qquad k=0,1,\ldots\,.
\end{equation}
Similarly,  all transverse vector harmonics are  of the form
\begin{equation}
 \label{eqn:vecthar}
 Y_\alpha^{A_1\ldots A_{k+1}}\eql Y_\alpha^{[(A_1}Y^{A_2}\ldots Y^{A_{k+1})]}\,,\qquad
 k=1,2,\ldots~,
\end{equation}
where the indices are symmetrized according to the $SO(6)$ hook
Young tableaux with $k$ boxes in the first row and one in the
second row. There are also longitudinal vector harmonics that are
obtained by differentiating the scalar harmonics. We will not need
them here since $\cA_{\alpha}$ is a transverse vector harmonic
(\ref{eqn:tran}).

Identities (\ref{eqn:derony}) turn all covariant differential
operators acting on harmonics into algebraic operations, while the
constraints  (\ref{eqn:algcon}) can be used to reduce products of
basic harmonics into  irreducible components. In particular,
following those steps, one obtains the familiar result for the
eigenvalues of the Laplacian used in \cite{Kim:1985ez}
\begin{equation}
 \label{laponschar}
 \boxxs Y^{A_1\ldots A_{k}}\eql -k(k+4)\,Y^{A_1\ldots A_k}\,,\qquad k=0,1,\ldots \,,
\end{equation}
\begin{equation}
 \label{laponvect}
 (\boxxs -4)\,Y_\alpha^{A_1\ldots A_{k+1}}\eql -(k+1)(k+3)\,Y_\alpha^{A_1\ldots A_{k+1}}\,,\qquad k=1,2,\ldots\,.
\end{equation}

The problem of solving  (\ref{eqn:laponvectors}-\ref{eqn:phifnct})
is now reduced to a finite dimensional linear algebra.  Suppose
that we start by choosing $\cA_\alpha$ as one of the vector
harmonics of order $k$, which is a polynomial of order $k+1$ in
the basic harmonics. Since the differentiation does not increase
that order, we conclude that $\Phi(\cA)$ is a polynomial of order
$2k+2$ or less  in the basic {\it scalar} harmonics. This shows
that $\Omega$ can be written as a finite sum
\begin{equation}
\Omega\eql \sum c_{A_1\ldots A_{2k+2}}Y^{(A_1}Y^{A_2}\ldots
Y^{A_{2k+2})}~,
\end{equation}
where the constant coefficients $c_{A_1\ldots A_{2k+2}}$ are then
 determined from the scalar equation.

Let us illustrate this using the already familiar case when
$\cA_\alpha$ is a  $SO(6)$ Killing vector. The latter are given by
the transverse  vector harmonics with $k=1$, which corresponds to
$n=2$ in (\ref{eqn:laponvectors}).  First consider
\begin{equation}
 \cA_\alpha\eql Y_\alpha^{AB}\eql Y_\alpha^AY^B-Y_\alpha^BY^A\,,
\end{equation}
for some fixed $A$ and $B$.  Using (\ref{eqn:algcon}) and
(\ref{eqn:derony})
 we obtain
 \begin{align}
 \cA_\alpha\cA^\alpha & \eql (Y^A)^2+(Y^B)^2\\
 & \eql Y^{AA}+Y^{BB}+{1\over 3}\,,
\end{align}
where the constant in the second line arises from subtracting
traces in the reduction to $SO(6)$ irreducible components.
Similarly,
\begin{equation}
 \cF_{\alpha\beta}\eql -2\left(Y_\alpha^A Y_\beta^B-Y_\beta^AY_\alpha^B\right)\,,
\end{equation}
and
\begin{equation}
 \cF_{\alpha\beta}\cF^{\alpha\beta}\eql 8\,\left[1-(Y^A)^2-(Y^B)^2\right]\,.
\end{equation}
Hence
\begin{align}
 \Phi(\cA)   \eql 4 \left[1-(Y^A)^2-(Y^B)^2\right]+4 \,\left[(Y^A)^2+(Y^B)^2\right]\eql 4\,.
\end{align}

More generally,   we take a linear combination of such harmonics,
\begin{equation}\label{eqn:Asqsq}
\cA_\alpha\eql{1\over
2}\sum_{A,B}\eta_{AB}\,Y_\alpha^{AB}\eql\sum_{A,B}\eta_{AB}\,Y_\alpha^AY^B\,,
\end{equation}
where $\eta_{AB}=-\eta_{BA}$. Then
\begin{equation}\label{AalphaAalpha}
 \cA_\alpha\cA^\alpha\eql \sum_{A,B,C}\eta_{AC}\eta_{BC}Y^{AB}+
\ds\frac{1}{3}\eta^2~, \qquad\qquad \Phi(\cA)\eql 4\eta^2\,,
\end{equation}
where
\begin{equation}\label{eqn:spherePhi}
\qquad \eta^2\eql\sum_{A<B}\eta_{AB}^2 \qquad \text{and} \qquad
Y^{AB} = Y^{A}Y^{B}-\ds\frac{1}{6}\delta^{AB}\,.
\end{equation}
Substituting this in the scalar equation (\ref{eqn:laponscalars})
with $n=2$, we get
\begin{equation}
 \boxxs\,\Omega+ 12\,\Omega\eql 4\eta^2\,.
 \end{equation}
The generic solution to this equation, which we discussed above,
is
 $ \Omega\eql \cA_\alpha\cA^\alpha  $. However, there is another obvious solution, which is simply the constant function
\begin{equation}\label{eqn:omconst}
\Omega\eql {1\over 3}\,\eta^2\,.
\end{equation}
From (\ref{AalphaAalpha}), the difference between the two
solutions is a sum of $k=2$ scalar harmonics, $Y^{A_1A_2}$, which
satisfy the homogenous equation (\ref{eqn:homogen}) (cf.
(\ref{laponschar})).

To summarize,  we have shown that there is a $15+20$ parameter
family\footnote{There is a 15 parameter degeneracy for the vector
harmonic with $k=1$ and a 20 parameter degeneracy for the scalar
harmonic with $n=2$. It is possible that some of these solutions
are equivalent.} of solutions on $S^5$
\begin{equation}
 \cA_\alpha\eql\cK_\alpha\,,\qquad \Omega\eql||\cK||^2+Y\,,
\end{equation}
where  $\cK_\alpha$ is a Killing vector and  $Y$ is a solution to
(\ref{eqn:laponscalars}) with $k=2$.

The solutions with $\Omega=\cK_\alpha\cK^\alpha$ and $Y=0$ arise
naturally from the null Melvin twist construction. If one takes
$\cK_\alpha$ to be a Killing vector along the Hopf fiber, its
length is constant and the solution reduces to (72) with constant
$\Omega$. However, since all Killing vectors on $S^5$ are
equivalent under the $SO(6)$ symmetry, the solution with constant
$\Omega$ should exist for any choice of $\cK_\alpha$, which indeed
is the case.

As a final example we consider a solution with a higher vector
harmonic. Let us take the harmonic \cite{PvanNreview:2008}
\begin{equation}
 Y_\alpha^{ABC}\eql 2 Y_\alpha^{ A}Y^{B }Y^C-Y_\alpha^BY^AY^C-Y_\alpha^CY^AY^B\,,
\end{equation}
with distinct $A$, $B$ and $C$. It has $k=2$, which gives the
dynamical exponent $n=3$. Similarly as above, we find
\begin{equation}
 \Phi(\cA)\eql 9\,\left[(Y^B)^2+(Y^C)^2\right]\,.
\end{equation}
However, unlike before, the mass term in (\ref{eqn:homogen}) is
equal to 32, which does not correspond to any of the eigenvalues
of the scalar Laplacian in   (\ref{laponschar}). Hence the
solution
\begin{equation}
 \Omega\eql {9\over 20}\,\left[(Y^B)^2+(Y^C)^2-{1\over 8}\right]\,,
\end{equation}
is unique, up to the degeneracy in the choice of vector harmonic
discussed above. This example illustrates explicitly the problem
we mentioned earlier that, for higher order harmonics, $\Omega$
may be negative in some region of $X_5$.

We conclude the discussion of $S^5$ with the observation that the
eigenvalues for the transverse vector harmonics in
(\ref{laponvect}) determine the dynamical exponents of our
solutions  to take  values
\begin{equation}
 n = k+1\,.
\end{equation}
Hence, we have here examples of scale-invariant, non-relativistic
type IIB solutions with a $B$-field and integer dynamical
exponents, $n\geq 2$.

\subsection{Vector harmonics on $T^{1,1}$}

We will not attempt to find explicit solutions with different
dynamical exponents generated by the vector harmonics on
$T^{1,1}$. Instead we will map out the relation between the
spectroscopy of vector eigenfunctions of the Laplacian on
$T^{1,1}$ and the dynamical exponents of the field theories dual
to the new gravity solutions.

The eigenfunctions of the vector Laplacian are labeled by weights
of the $SU(2)\times SU(2)\times U(1)$ isometry group of $T^{1,1}$
\cite{Ceresole:1999zs}, \cite{Ceresole:1999ht},
\begin{equation}
 \nabla^2 \mathcal{A}_\alpha^{l_1,l_2,l_3} = \lambda(l_1,l_2,l_3)
 \mathcal{A}_\alpha^{l_1,l_2,l_3}~.
\end{equation}
There are four series of eigenvalues
\begin{eqnarray}
\lambda_{1,2} &\eql& 3 + h(l_1,l_2,l_3\pm2)~,\\
\lambda_{3,4} &\eql& h+4 \pm 2\ds\sqrt{h+4}~,
\end{eqnarray}
where $h(l_1,l_2,l_3)$ are the scalar eigenvalues of the Laplacian
on $T^{1,1}$
\begin{equation}
h(l_1,l_2,l_3) \eql 6 \left(l_1(l_1+1)+l_2(l_2+1) -
\ds\frac{l_3^2}{8}\right)\,.
\end{equation}
Here $l_1,l_2$ could be integers or half-integers and $l_3$ is an
integer. The values of the dynamical exponents of the
gravitational backgrounds are determined by the solutions to any
of the four algebraic equations
\begin{equation}
n(n+2) \eql 4- \lambda_{i}(l_1,l_2,l_3) ~, \qquad\qquad i =
1,2,3,4\,.
\end{equation}
It follows that for $T^{1,1}$ as the internal manifold, the
dynamical exponents of the non-relativistic solutions are not
arbitrary integers, as is the case for $S^5$. In fact, for generic
values of $(l_1,l_2,l_3)$ the dynamical exponent, $n$, is
irrational.

\section{Conclusions}

We have found a large class of supersymmetric IIB solutions with
non-vanishing $B$-field which are invariant under the
Schr\"{o}dinger symmetry. The solutions are obtained by the null
Melvin twist from supersymmetric type IIB solutions of the form
$AdS_5\times X_5$. The field theories dual to these solutions form
a very special class of non-relativistic field theories. They can
be obtained from the relativistic $\mathcal{N}=1$ superconformal
Yang-Mills theories dual to $AdS_5\times X_5$ by performing a
discrete light-cone quantization accompanied by a twist. The twist
amounts to modifying all products of chiral superfields in the
Lagrangian of the relativistic theory by phases proportional to
the charges of the fields under the $U(1)$ global symmetries used
in the null Melvin twist. This class of field theories was
discussed in \cite{Bergman:2000cw,Dasgupta:2000ry,Bergman:2001rw},
see also \cite{Adams:2008wt}.

We also found a quite general class of type IIB solutions with
dynamical exponents different from two and non-vanishing
$B$-field. As discussed in Section 5, the $B$-field is determined
by a vector harmonic on an Einstein manifold, $X_5$. The metric is
obtained by solving an inhomogeneous scalar Laplace equation on
$X_5$. The solutions are invariant under the Galilean group and
dilatations, but are not invariant under special conformal
transformations, and thus break the full Schr\"{o}dinger symmetry.
The dual field theories should be scale invariant and invariant
under Galilean transformations. Since we did not generate the
gravity solutions by some twist of known relativistic solutions
with clear D-brane interpretation, the detailed structure of the
dual non-relativistic field theories is not clear at present. It
would be very interesting to reduce the solutions of Section 5 to
five dimensions and to see whether they can be obtained as
solutions to five-dimensional gravity coupled to some massive
fields \cite{Son:2008ye,Balasubramanian:2008dm,Maldacena:2008wh}.
More generally, it is important to understand compactifications
and consistent truncations of type IIB supergravity with massive
fields \cite{Maldacena:2008wh,Gauntlett:2009zw}.

There are several directions in which our analysis could be
extended. It is interesting to find eleven-dimensional analogs of
our solutions. Some supersymmetric non-relativistic solutions of
the eleven-dimensional supergravity are known, \cite{Donos:2009en,
Colgain:2009wm}. Perhaps it is possible to find more general
classes of such solutions using the results of this paper as a
guide. A natural way to proceed is to start with a Freund-Rubin
solution of the form $AdS_4\times X_7$ where $X_7$ is a
seven-dimensional Sasaki-Einstein manifold with $U(1)^4$
isometry.\footnote{There exists an infinite family of such
manifolds with explicitly known metrics
\cite{Cvetic:2005ft,Gauntlett:2004hh}.} Then one can reduce the
solution along the $U(1)$ R-symmetry to get a solution of IIA
supergravity with $U(1)^3$ symmetry. The generalized null Melvin
twist applied to this solution would generate a non-relativistic
type IIA solution with a $Sch_4$ non-compact space-time. This
solution could be uplifted to a solution of eleven-dimensional
supergravity. It is natural to expect that for the null Melvin
twist performed along $U(1)$ isometries that leave the $X_7$
Killing spinor invariant, the twisted eleven-dimensional
background will preserve some supersymmetry. It should also be
possible to use this eleven-dimensional solution as a guide for
constructing more general solutions along the lines of Section~5.

There are supersymmetric extensions of the Schr\"{o}dinger algebra
with various amounts of supersymmetry \cite{Sakaguchi:2008rx}. It
is interesting to explore the connection between these
superalgebras and the supersymmetric Schr\"{o}dinger invariant IIB
solutions found here and also those in
\cite{Hartnoll:2008rs,Donos:2009en}. It is natural to explore
whether any of the supersymmetric type IIB solutions presented
here could be realized as non-relativistic supercosets along the
lines of \cite{SchaferNameki:2009xr}.

We would like to note that it is straightforward to find finite
temperature counterparts of all solutions discussed in Section 2
and Section 4. One should start with a $AdS_5\times X_5$ black
hole solution and apply the generalized null Melvin twist. This
was done in \cite{Bobev:2009zf} for the $AdS_5\times S^5$ black
hole. It would be also interesting to construct the finite
temperature versions of the solutions with general dynamical
exponents found in Section 5 and explore their thermodynamics.

Finally it is tempting to speculate that there might be
supersymmetric (and non-supersymmetric) non-relativistic analogs
of the familiar RG flow solutions of the type IIB and
eleven-dimensional supergravities. A modest attempt to construct
such solutions was made in \cite{Bobev:2009zf}, but there is
certainly much more to explore.

\bigskip
\textbf{\large Acknowledgments}
\bigskip

We are grateful to Veronika Hubeny, Mukund Rangamani and Nick
Warner for many useful conversations and patient explanations. We
would also like to thank Tameem Albash, St\'{e}phane Detournay,
Clifford Johnson, Mike Mulligan and Kentaroh Yoshida for useful
discussions and comments.  This work is supported in part by the
Department of Energy grant DE-FG03-84ER-40168. NB is supported
also by a Graduate Fellowship from KITP and in part by the
National Science Foundation under Grant No.\ PHY05-51164.

\renewcommand{\theequation}{A.\arabic{equation}}
\setcounter{equation}{0}  
\section*{Appendix A. Generalized null Melvin twist}  

The null Melvin twist \cite{Gimon:2003xk,Hubeny:2005qu} (see, also
\cite{Herzog:2008wg,Adams:2008wt,Bobev:2009zf}) is a solution
generating technique  which can be used to construct explicitly
new solutions of the type IIB supergravity starting  from  a known
solution with at least $U(1)$ isometry. In this Appendix we
summarize the main steps of the construction as applied recently
in  \cite{Bobev:2009zf} to  solutions with $U(1)^3$ isometry.  We
show that by examining explicit formulae for the twisted solutions
in \cite{Bobev:2009zf}, one is naturally led to rewrite the result
of the null Melvin twist along any isometry given by a Killing
vector, $\K$, in terms of intrinsic quantities without reference
to any specific coordinates.

Consider a type IIB solution of the form $AdS_5\times X_5$, where
$X_5$ is an Einstein manifold with $U(1)^3$ isometry. For
definiteness, let us take
\begin{equation}
ds^2_{10} = \ds\frac{1}{z^2} (-dt^2+dy^2+dx_1^2+dx_2^2+dz^2) \\+
ds^2_{X_5}\,,
\end{equation}
\begin{equation}
F_{(5)} = (1+\star) ~\text{vol}_{AdS_5}\,,
\end{equation}
where we have  normalized the $AdS_5$ and $X_5$ to be of unit
radius. In all cases of interest, the metric on the internal
manifold $X_5$ can be written, at least locally, in the form
\begin{equation}\label{eqn:intmet}
ds^2_{X_5} = f_1 d\theta_1^2 + f_2d\theta_2^2 + \sum_{i,j=1}^3
f_{ij} d\phi_id\phi_j\,,
\end{equation}
where all functions $f_i$ and $f_{ij}$ depend only on $\theta_1$
and $\theta_2$. The angles $\phi_1,\phi_2,\phi_3$ parametrize
directions along the three $U(1)$'s.

The generalized null Melvin twist
\cite{Gimon:2003xk,Hubeny:2005qu,Bobev:2009zf} consists of the
following operations that are  straightforward to implement:
\begin{itemize}
\item a boost in the $(t,y)$ plane with parameter $\gamma_0$,
\item a T-duality along $y$,
\item a shift of all three $U(1)$ isometries of $X^5$ given by $\phi_i \to
\phi_i + a_i y$,
\item another T-duality along $y$,
\item an inverse boost in the $(t,y)$ plane with
parameter $-\gamma_0$,
\item a limit $a_i\to 0$, $\gamma_0\to
\infty$ such that $\eta_i\equiv a_i
\cosh\gamma_0=a_i\sinh\gamma_0$ remain finite.
\end{itemize}
Introducing  light-cone coordinates
\begin{equation}
 u\eql t+y\,,\qquad v\eql {1\over 2}(t-y)\,,
\end{equation}
the resulting spacetime  is a product space $Sch_5\times X_5$,
with the metric
\begin{equation}\label{eqt: genSE}
ds^2_{10} \eql - \ds\frac{\Omega}{z^4}~ du^2+\ds\frac{1}{z^2}
(-2dudv+dx_1^2+dx_2^2+dz^2)\\+ ds^2_{X_5}\,,
\end{equation}
a nontrivial two-form potential
\begin{equation}
B_{(2)} = \ds\frac{1}{2z^2} [\partial_{\eta_1}\Omega ~d\phi_1 +
\partial_{\eta_2}\Omega ~d\phi_2 + \partial_{\eta_3}\Omega ~d\phi_3]\wedge
du\,,
\end{equation}
and the five-form flux
\begin{equation}\label{eqn:f5flux}
  F_{(5)}\eql (1+\star){\rm vol}_{Sch_5}\,,
  \end{equation}
where $\Omega$ is explicitly given by
\begin{equation}\label{eqn:omega}
\Omega (\theta_1,\theta_2) =
\sum_{i,j=1}^3f_{ij}\,\eta_i\eta_j\,.\end{equation}
Note that since ${\rm vol}_{AdS_5}\eql {\rm vol}_{Sch_5}$, the
five-form flux is in fact invariant under the twist.

The entire twisted solution is completely determined by $\Omega$
in (\ref{eqn:omega}). In order that the metric in (\ref{eqt:
genSE}) is well defined over the entire space, $Sch_5\times X_5$,
this $\Omega$ must be a scalar function on $X_5$. This in turn
implies that the twist parameters $\eta_i$'s should be viewed as
coordinates of a vector field on $X_5$, rather than constants.
Indeed, the solution (\ref{eqt: genSE}-\ref{eqn:f5flux}) can be
easily recast into a form that makes this manifest.

The Killing vectors of the $U(1)^3$ isometry for the  internal
metric (\ref{eqn:intmet}) are linear combinations of the three
Killing vectors
\begin{equation}
\K_{(i)} = \ds\frac{\partial}{\partial \phi_i}\,,\qquad i=1,2,3\,.
\end{equation}
We observe that in terms of the Killing vector
\begin{equation}
\K \eql \sum_{i=1}^3 \eta_i\,\K_{(i)}\eql  \ds\sum_{i=1}^{3}
\eta_i \,\frac{\partial}{\partial \phi_i}\,,
\end{equation}
we have simply
\begin{equation}\label{eqn:newomega}
\Omega=||\K||^2\,,
\end{equation}
where $||\K||$ is the norm of $\K$. Furthermore,
\begin{equation}\label{eqn:newb2}
  B_{(2)}\eql {1\over
z^2}\,\kappa\wedge du\,,
\end{equation}
where  $\kappa$ is the one form dual to Killing vector, $\K$, with
respect to the internal metric.

{Eqs.\ (\ref{eqn:newomega}) and (\ref{eqn:newb2})  show that the
twisted solution is well defined over the entire internal
manifold. In particular, one can use them to write down the
solution (\ref{eqt: genSE})-(\ref{eqn:omega}) in   terms of
arbitrary coordinates  ($\xi^\alpha$) on $X_5$,
\begin{align}
 ds_{X_5}^2  \eql
g_{\alpha\beta}\,d\xi^\alpha  d\xi^\beta\,,  \label{eqn:ricci}
 \qquad R_{\alpha\beta}  \eql 4\,g_{\alpha\beta}\,,
\end{align}
and
\begin{equation}
\label{eqn:omandk}
  \Omega=
g_{\alpha\beta}\,\K^\alpha \K^\beta\,,\qquad
 \kappa_\alpha=g_{\alpha\beta}\,\K^\beta\,.
\end{equation}
Since the metric is block diagonal, (\ref{eqn:omandk}) has the
same form when written  in terms of ten-dimensional coordinates.

The explicit background fields in  (\ref{eqt:
genSE})-(\ref{eqn:omega}) were obtained by applying the null
Melvin twist along an arbitrary  Killing vector. Therefore, by
construction one is guaranteed to get a solution of the type IIB
supergravity. However, it is also illuminating to verify this
explicitly starting with  an Ansatz for the  fields as in
(\ref{eqt: genSE}-\ref{eqn:f5flux}),    (\ref{eqn:newomega}) and
(\ref{eqn:newb2}), where $X_5$ is an arbitrary Einstein manifold
with a  globally defined vector field $\K$.  Using the formulae
for the spin connection and the fluxes in Section 3, we find that
the Maxwell and the Einstein equations reduce to two equations for
$\K$,
\begin{equation}\label{eqn:max}
  \boxx \,\K_\alpha+4\,\K_\alpha\eql 0\,,
\end{equation}
and
\begin{equation}\label{eqn:redein}
 \K^\alpha\boxx\,\K_\alpha+{1\over 2}\left(\nabla_\alpha\K_\beta+\nabla_\beta\K_\alpha\right)\nabla^\alpha\K^\beta+4\,\K_\alpha\K^\alpha\eql 0\,,
 \end{equation}
respectively, where $\boxx=\nabla^\alpha\nabla_\alpha$ is the
covariant Laplacian on $X_5$. The two equations imply that
$\nabla_{(\alpha}\K_{\beta)}=0$, as the second term in
(\ref{eqn:redein}) is manifestly positive. Thus $\K^\alpha$ must
be a Killing vector and (\ref{eqn:redein}) follows from
(\ref{eqn:max}). It is a standard fact that Killing vectors are
eigenfunctions  of  the Laplacian on an Einstein manifold so that
the latter equation  is always satisfied. Indeed, we have
\begin{align}\label{eqn:killderLap}
 \nabla^\alpha(\nabla_\alpha\K_\beta+\nabla_\beta\K_\alpha) & \eql
 \boxx\,\K_\beta+R^\alpha{}_\beta\K_\alpha\\[6 pt]
\notag & \eql\boxx\,\K_\beta +4\,\K_\beta\,,
\end{align}
where we used  that $\nabla^\alpha\K_\alpha=0$. Note that the
normalization of the mass term in (\ref{eqn:max}) corresponds to
the unit radius of $X_5$ in (\ref{eqn:ricci}).



\end{document}